\documentclass[preprint,aps]{revtex4}
\usepackage{graphicx}
\def\be{\begin{equation}}
\def\ee{\end{equation}}
\def\bea{\begin{eqnarray}}
\def\eea{\end{eqnarray}}

\def\a{\alpha}

\def\e{\epsilon}

\begin{document}
\title{Exact solitary wave solutions of the nonlinear Schr\"odinger
equation with a source}
\author{T. Soloman Raju, $^1$ \thanks{panisprs@prl.ernet.in}
C. Nagaraja Kumar, $^2$ \thanks{cnkumar@mail.pu.ac.in} and
Prasanta K. Panigrahi, $^{1,3}$ \thanks{prasanta@prl.ernet.in}}
\address{$^1$ School of Physics, University of Hyderabad, Hyderabad, 500 046, India\\
$^2$  Physics Department, Panjab University, Chandigarh, 160 014, India\\
$^3$ Physical Research Laboratory, Navrangpura, Ahmedabad, 380
009, India}

\begin{abstract}
We use a fractional transformation to connect the traveling wave
solutions of the nonlinear Schr\"odinger equation (NLSE),
phase-locked with a source, to the elliptic functions satisfying,
$f^{\prime\prime}\pm af\pm \lambda f^{3}=0$. The solutions are
{\it necessarily} of the rational form, containing both
trigonometric and hyperbolic types as special cases. Bright, and
dark solitons,  as also singular solitons, are obtained in
suitable range of parameter values.
\end{abstract}
\maketitle

Much attention has been paid to the study of the externally driven
NLSE, after the seminal work of Kaup and Newell \cite{kaup}. This
equation features prominently in the problem of optical pulse
propagation in asymmetric, twin-core optical fibers
\cite{gil,synder,boris}; currently an area of active research. Of
the several applications of an externally driven NLSE, perhaps the
most important ones are to long Josephson junctions \cite{sam},
charge density waves \cite{newell}, and the plasmas driven by rf
fields \cite{bekki}. The phenomenon of autoresonance \cite{land},
indicating a continuous phase locking between the solutions of
NLSE and the driving field has been found to be a key
characteristic of this system. In the presence of damping, this
dynamical system exhibits rich structure including bifurcation.
This is evident from analyses, around a constant background, as
well as numerical investigations
\cite{smirnov,barashenkov,nistazakis}. Although the NLSE is a
well-studied integrable system \cite{agrawal}, no exact solutions
have so far been found for the NLSE with a source, to the best of
the authors' knowledge. All the above inferences have been drawn
through perturbations around solitons and numerical techniques.

In this Letter, we map {\it exactly}, the traveling wave solutions
of the NLSE phase-locked with a source, to the elliptic functions,
through a fractional transformation (FT).  It was found that the
solutions are {\it necessarily} of the rational type, with both
the numerator and denominator containing terms quadratic in
elliptic functions, in addition to having constant terms. It is
well-known that the solitary wave solutions of the NLSE
\cite{whitham,novikov} are cnoidal waves, which contain the
localized soliton solutions in the limit, when the modulus
parameter equals one \cite{abramowitz}. Hence, the solutions found
here, for the NLSE with a source, are nonperturbative in nature.
We find both bright and dark solitons as also singular ones.
Solitons and solitary pulses show distinct behavior. In the case,
when the source and the solutions are not phase matched,
perturbation around these solutions may provide a better starting
point.

For nonlinear equations, a number of transformations are
well-known in the literature, which map the solutions of a given
equation to the other \cite{drazin,das}. The familiar example
being the Miura  transformation \cite{miura}, which maps the
solutions of the modified KdV  to those of the KdV equation. To
find static and propagating solutions, appropriate transformations
have also been cleverly employed, to connect the nonlinear
equations to the ones satisfied by the elliptic functions:
$f^{\prime\prime}\pm af\pm \lambda f^{3}=0$. Here and henceforth,
prime denotes derivative with respect to the argument of the
function. Solitons and solitary wave solutions of KdV, NLSE, and
sine-Gordon etc., can be easily obtained in terms of the elliptic
functions in this manner.

We consider the NLSE with an external traveling wave source:
\begin{equation}
i\frac{\partial \psi}{\partial t} + \frac{\partial^{2}
\psi}{\partial x^2} +g \mid \psi \mid^2 \psi +\mu \psi=k
e^{i[\chi(\xi) - \omega t ]}\quad ,
\end{equation}
where $g$, $\mu$, and $k$ are real and $\xi=\a(x-vt)$. The
traveling wave solution, phase-locked with the external source is
taken to be $ \psi ( x , t ) = e^{i[\chi(\xi) - \omega
t]}\rho(\xi). $  Separating the real and imaginary parts of Eq.
(1), and  integrating the imaginary part one gets,
\begin{equation}
\chi^\prime = \frac{v}{2\alpha}+\frac{c}{\a\rho^2},
\end{equation}
where $c$ is the integration constant. In order that, the external
phase is independent of $\psi$, we put $c=0$ to obtain,
\begin{equation}
 \alpha^2 \rho^{\prime\prime} +g\rho^3-\epsilon \rho - k =
 0,
\end{equation}
where $\epsilon=\omega-\frac{v^2}{4}-\mu$. For illustrating our
procedure, we assume that $\e
> 0$, and $g > 0$, the other cases can be studied analogously.
After a suitable scaling of the field variable, the above equation
can be cast into the convenient form,
\begin{equation}
\rho^{\prime\prime}+\rho^3-\rho-k=0.
\end{equation}

We find, after a straightforward but lengthy algebra that, the
following FT, \be \rho(\xi)=\frac{A+B f^\delta(\xi,m)}{1+D
f^\delta(\xi,m)}, \ee
 for $AB-D\neq
0$, maps the solutions of Eq. (4) to the elliptic functions
$f(\xi)$, provided $\delta=2$. $f(\xi)$ satisfies
$f^{\prime\prime}= f-  f^{3}$, with a conserved quantity
$E_{0}=f^{{\prime}2}/2+(1/4)f^{4}-f^{2}/2$.  The above equation
has four bounded periodic solutions which are Jacobi elliptic
functions: ${\rm cn}(\xi,m)$, ${\rm sd}(\xi,m)$, ${\rm
dn}(\xi,m)$, and ${\rm nd}(\xi,m)$, where $m$ is the modulus
parameter. Apart from finite energy oscillatory solutions, the
above has localized soliton solutions. It should be noted that for
the attractive case $(g<0)$, the bounded solutions are ${\rm
sn}(\xi,m)$, and ${\rm cd}(\xi,m)$. For $AD=B$, one only finds a
constant amplitude solution.

Before elaborating on specific cases, a number of interesting
features emerging from the above mapping is worth mentioning.
First of all, the nontrivial solutions are necessarily of the
rational type, i.e., $D\neq 0$, in the presence of the source.
Secondly, $E_{0}=0$, and  $E_{0}\neq 0$ cases, show
characteristically different behavior. For example, for $E_{0}\neq
0$ case, one can have solutions with $A=0$, and $B\neq 0$, of the
type \be \rho(\xi)=\frac{(k/4E_{0})f^2}{1+(1/8E_{0})f^2}. \ee
 However,
$A\neq 0$, and $B=0$ is not allowed. For $E_{0}=0$ case, one can
have for $B=0$, and $A\neq0$, a singular solution of the following
form, \be \rho(\xi)=\frac{2k}{1-f^2}. \ee As we will see
explicitly below, nonsingular solutions are also possible. Like
the former case,  here $B\neq0$, and $A=0$ is not allowed in the
presence of the source.

For explicitness, we consider the unscaled equation containing all
the parameters and illustrate below various type of solutions,
taking $f=cn(\xi,m)$.  Other cases can be similarly worked out.
The consistency conditions are given by,
 \bea
A\epsilon-2\alpha^{2}(AD-B)(1-m)+gA^{3}-k&=0,\\\nonumber \\
 2\epsilon AD+\epsilon
 B+6\alpha^{2}(AD-B)D(1-m)- \nonumber \\
 4\alpha^{2}(AD-B)(2m-1)+3gA^{2}B-3kD&=&0,\\\nonumber\\
A\e D^{2}+2\e BD+4\alpha^{2}(AD-B)D(2m-1)+ \nonumber \\
6\a^{2}(AD-B)m+3gAB^{2}-3kD^{2}&=&0,\\\nonumber\\
\e BD^{2}-2\a^{2}(AD-B)Dm+gB^{3}-kD^{3}&=&0.
 \eea

The above equations clearly indicate that the solutions, for
$m=1$, $m=0$ and other values of $m$, have distinct properties.
For example, when $m=1$, $A$ is obtained as the solution of the
cubic equation [Eq. (8)], containing the source strength $k$.
Similarly, for $m=0$, either $B$, or $D$ appears as the solution
of Eq. (11). As noted earlier, when $D=0$, $B$ also equals zero,
indicating only a constant solution $(A\neq 0)$. One needs to be
careful in choosing the real solutions of the above cubic
equations, for suitable range of the parameter values. Although a
wide class of solutions are allowed; for brevity, we only outline
a few of the interesting solutions and their properties in a
restricted range of equation parameters.

We start with the localized solitons, by taking $m=1$ $({\rm
cn}(\xi,1)={\rm sech}(\xi) )$, with the parameter values $A=1$,
$B=\delta$, and $D=1-\delta$ we obtain
 \be
\rho(\xi)=\frac{1+\delta {\rm sech}^2(\xi)}{1+\Gamma {\rm
sech}^2(\xi)};
 \ee
here the amplitude, width, and velocity are related as,
$\delta=-(Q+\sqrt{Q^2 -4PR})/2Q$, with $Q=\e-2\a^2-3k$,
$P=2\e-3k$, $R=-k-2\a^2$, and $\Gamma=1+\delta$. Here the width
$\a$ is the only independent parameter. As the above form of the
solution indicates, both nonsingular and singular solitons are
possible solutions depending on the values of $\e$, and the source
strength $k$. We have checked that periodic solutions are also
allowed. Below we give a few more solutions, illustrating the
differences between the localized soliton  and the periodic
solitary wave solutions.

Case(I): Trigonometric solution.- In the limit $m=0$,  unlike the
unperturbed NLSE, in this case, one finds rational solutions of
the trigonometric type. Apart from the general solutions,
interestingly, for these, one can obtain special cases, where
$A=0$ and $B\neq 0$ is allowed. However, the vice-versa is
forbidden.  Following is an example of this type of nonsingular
periodic solutions, for the repulsive case:

\begin{equation}
\rho(\xi)=(-\frac{2k}{\epsilon})\frac{\cos^2 (\xi)} { 1 -
\frac{2}{3} \cos^2 (\xi )}.
\end{equation}
Here, $\e$ has to be negative since, $\alpha^2=-\epsilon/4$; and
is given by $\epsilon=(-27gk^2/2)^{1/3}$. This periodic solution
is found to be stable, as evidenced from the numerical simulations
seen in Fig.1.

Case(II): Hyperbolic solution.- Unlike the above periodic case,
here one finds that, the solutions with $B=0$, and $A\neq 0$ are
allowed, the vice-versa  not being true. Hence, these localized
solitons behave differently from the soliton trains. There are
both singular and nonsingular solutions. For $B=0$, and $m=1$, we
found that $\alpha^2=\epsilon/4$, and
$\epsilon=(-27gk^2/2)^{1/3}$. This yields, the singular hyperbolic
solution,
\begin{equation}
\rho(\xi)=(\frac{3k}{\epsilon})\frac{1}{1-\frac{3}{2}{\rm sech}^2
(\xi)}.
\end{equation}
The singularity here may correspond to extreme increase of the
field amplitude due to self-focussing, as is known for the other
nonlinear systems. We have also found nonsingular solutions of the
above type.

Case(III): Pure cnoidal solutions.-
In the general case, for
$0<m<1$, from Eq. (11), one obtains the width $\a$, in terms of
the other parameters. The other three equations then yield $A$,
$B$, and $D$; in the process one encounters a cubic equation whose
real roots need only to be considered. This puts restriction on
the solution parameters. Interestingly, the constraints imply an
$m$ independent condition on the parameters in the form
 \bea \nonumber
 &&5\e AD^2+\e AD^3+\e BD+5\e BD^2+3gA^2
BD\\
&&+3gADB^2 -6kD^2-6kD^3
 +3gA^3 D^2 +3gB^3=0 \,.\nonumber
 \eea
 Below we list a few special cases.
 For $A=0$, $D=1$,  and $m=5/8$; we found
that for $\alpha^2=(2/7)\epsilon$, and
$\epsilon=7(-gk^2/18)^{1/3}$; this corresponds to attractive case.
Explicitly, the solution is given by,
\begin{equation}
\rho(\xi)=(\frac{14k}{3\epsilon})\frac{{\rm cn}^2 (\xi,m)}{1+{\rm
cn}^2 (\xi,m)}.
\end{equation}

For $A=0$, and $m=1/2$; it is found that
$\alpha^2=\epsilon/2\sqrt{3}$ and
$\epsilon=(-27gk^2)^{\frac{1}{3}}$, for which
\begin{equation}
\rho(\xi)=(\frac{2\sqrt{3}k}{\epsilon})\frac{{\rm cn}^2
(\xi,m)}{1+\frac{1}{\sqrt{3}}{\rm cn}^2 (\xi,m)}.
\end{equation}

\begin{figure}
\centering
\includegraphics[width=3.75in]{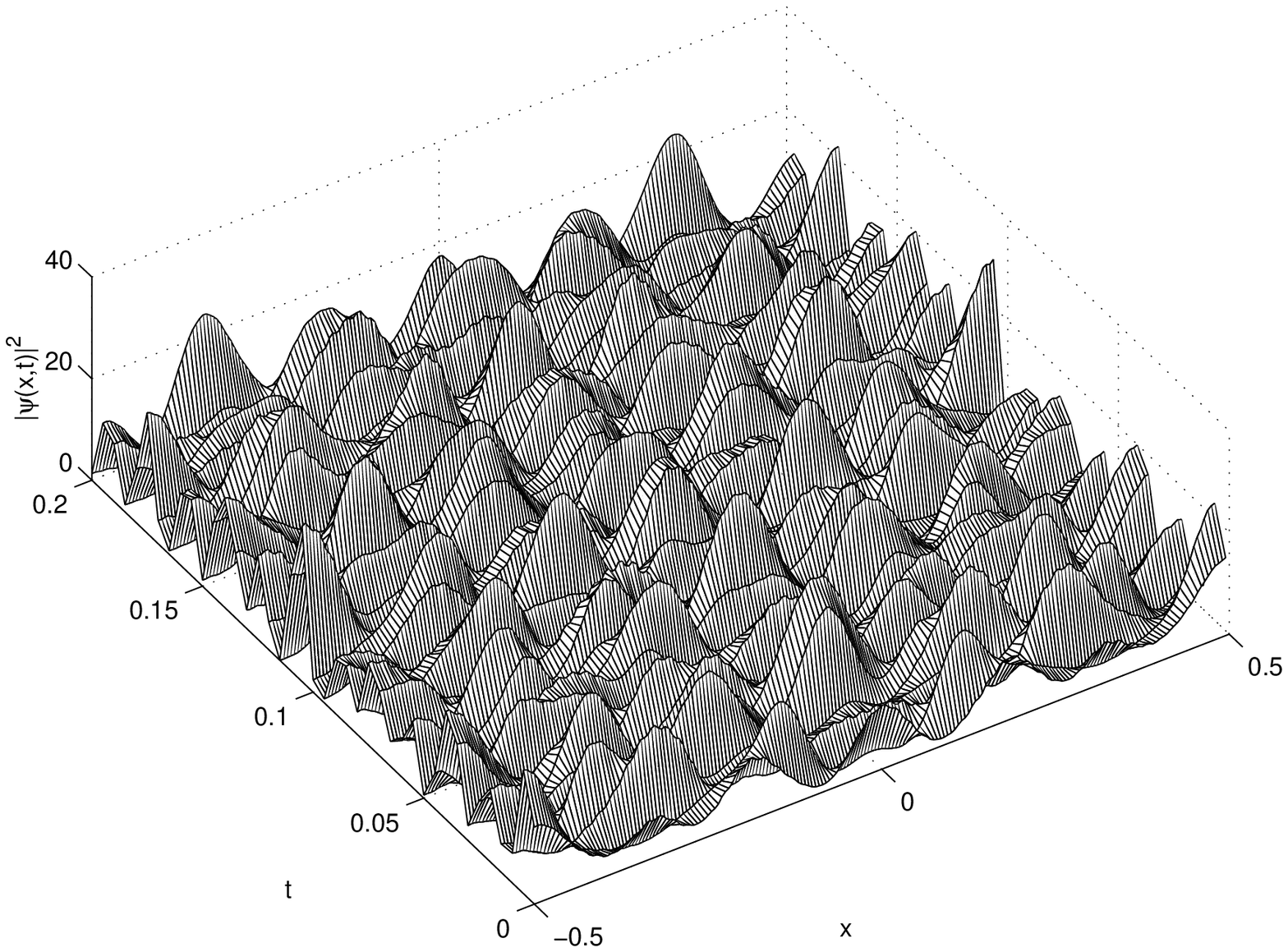}
\includegraphics[width=3.75in]{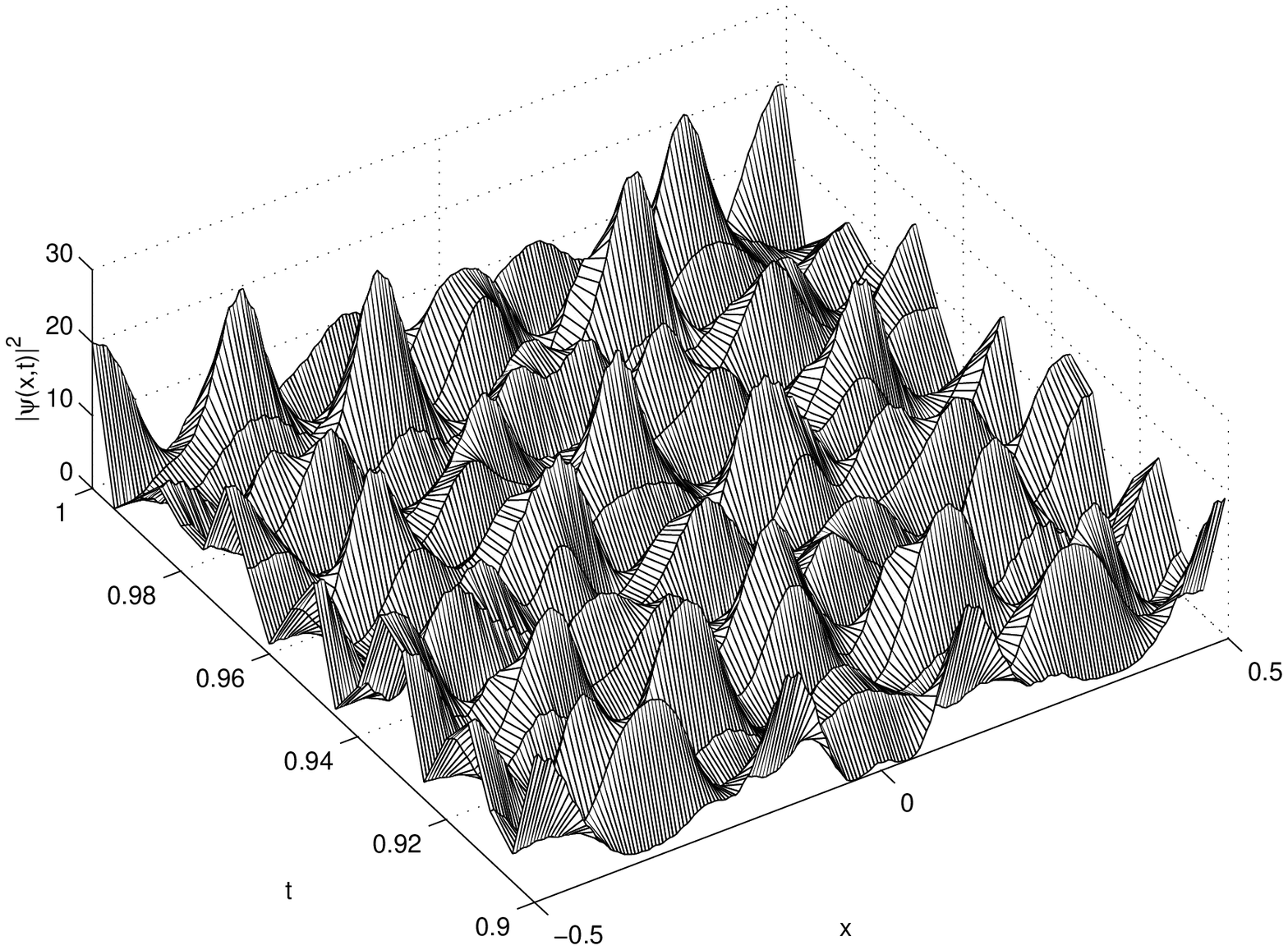}
 \caption{Plot depicting the evolution of the
 trigonometric solution, for various times.}
\end{figure}
Since the localized solitons are usually robust, we have performed
numerical simulations to check the stability of the solutions
pertaining to Case(I), {\it i.e.,} the trigonometric solution. It
is worth pointing out that the
 numerical techniques based on the fast Fourier transform (FFT) are expensive as
 they require the  FFT of the external source. Hence, we have used the
 Crank-Nicholson finite difference method \cite{press}
to solve the NLSE with a source, which is
 quite handy, and unconditionally stable.
The initial conditions chosen from the exact solution are knitted
on a lattice with a grid size $dx=0.005$, and $dt=5.0\times
10^{-6}$. The evolution of this solution, as depicted in Fig.1
indicate that it is a stable solution.

In conclusion, we have used a fractional transformation to connect
the solutions of the phase-locked NLSE with the elliptic
functions, in an exact manner. The solutions are necessarily of
the rational type that contain solitons, solitary waves, as also
singular ones. Our procedure is applicable, both for the
attractive and repulsive cases. Because of their exact nature,
these will provide a better starting point for the treatment of
general externally driven NLSE. Considering the utility of this
equation in fiber optics and other branches of physics, these
solutions may find practical applications.

 We acknowledge profitable discussions with Dr. E. Alam regarding
 the algorithm that has been implemented here, and Prof. A. Khare
 for many useful discussions.


\begin{references}

 \bibitem{kaup} D.J. Kaup and A.C. Newell, Proc. R. Soc. London, Ser. A
     {\bf 361}, 413 (1978).

\bibitem{synder} A.W. Snyder and J.D. Love, {\it Optical Waveguide Theory}
(Chapman and Hall, London, 1983).

\bibitem{boris} B.A. Malomed, Phys. Rev. E {\bf 51}, R864 (1995).

\bibitem{gil} G. Cohen, Phys. Rev. E {\bf 61}, 874 (2000).

 \bibitem{sam} P.S. Lomdahl and M.R. Samuelsen, Phys. Rev. A {\bf 34}, 664
(1986).

\bibitem{newell} D.J. Kaup and A.C. Newell, Phys. Rev. B {\bf 18}, 5162 (1978).

\bibitem{bekki} K. Nozaki and N. Bekki, Physica D {\bf 21}, 381 (1986).

\bibitem{land} L. Friedland, Phys. Rev. E {\bf 58}, 3865 (1998).

\bibitem{smirnov} I.V. Barashenkov, Yu. S. Smirnov and N.V. Alexeeva, Phys.
Rev. E {\bf 57}, 2350 (1998).

\bibitem{barashenkov} I.V. Barashenkov,  E.V. Zemlyanaya, and M. B\"ar, Phys.
Rev. E {\bf 64}, 016603 (2001).

\bibitem{nistazakis} H.E. Nistazakis, P.G. Kevrekidis, B.A. Malomed,
D.J. Frantzeskakis, and A.R. Bishop Phys. Rev. E {\bf 66}, R015601
(2002).

\bibitem{agrawal} G.P. Agrawal, {\it Nonlinear Fiber optics} (Academic Press,
Boston, 1989).

\bibitem{whitham} G.B. Whitham, {\it Linear and Nonlinear Waves}
(Wiley, New York, 1974).

\bibitem{novikov} S.P. Novikov, S. V. Manakov, L. P. Pitaevsky,
and V.E. Zakharov, {\it Theory of Solitons. The Inverse Scattering
Method} (Consultants Bureau, New York, 1984) and references
therein.


\bibitem{abramowitz} M. Abramowitz  and A. Stegun, {\it Handbook of Mathematical
Functions} (Dover Publications, New York, 1970).

\bibitem{das} A. Das, {\it Integrable Models} (World Scientific, Singapore,
1989).

\bibitem{drazin} P.G. Drazin and R.S. Johnson, {\it Solitons: An
introduction} (Cambridge University Press, Cambridge, 1989).


\bibitem{miura} R.M. Miura, J. Math. Phys. {\bf 9}, 1202 (1968).


\bibitem{press} W.H. Press, B.P. Flannery, S.A. Teukolsky and W.T.
Vellerling, {\it Numerical Recipes in Fortran} (Cambridge
University Press, Cambridge, 1992).

\end{references}
\end{document}